# ZZ CETI STARS: FRACTAL ANALOGUES OF EXCITED He$^+$ IONS?

**Robert L. Oldershaw**

Earth Sciences Building

Amherst College

Amherst, MA 01002

rloldershaw@amherst.edu

**Running Head:** ZZ CETI STARS AND HELIUM IONS

**Abstract:** Multi-periodic pulsation phenomena of ZZ Ceti variable stars are analyzed within the context of a fractal cosmological paradigm that emphasizes discrete self-similarity in nature. The majority of ZZ Ceti stars are found to be analogues of highly perturbed $^4$He$^+$ ions oscillating at transition periods characterized by $1 \leq n \leq 8$, $1 \leq \Delta n \leq 6$ and $1 \leq l \leq 7$. A quantitative test comparing relevant stellar and atomic periods provides preliminary support for the analysis. The low amplitudes and multiple periods of the ZZ Ceti stars argue that their pulsations take place below the threshold needed for discrete energy level transitions. This paper completes a 3-part series that explores the possibility of discrete cosmological self-similarity in variable stars; the first and second papers discussed RR Lyrae and d Scuti stars, respectively.

1. INTRODUCTION

This paper is the last segment of a 3-part study of discrete self-similar relationships between well-defined classes of variable stars and their specific Atomic Scale analogues. The *first part*, entitled "Discrete Self-Similarity Between RR Lyrae Stars and Their Constituent Helium Atoms," demonstrated that RR Lyrae variables are the Stellar Scale equivalents of $^4$He atoms undergoing single-level transitions between excited states with $7 \leq n \leq 10$ and $l \leq 1$. A high resolution sample of RR Lyrae oscillation periods was used to document strong correlations between observed Stellar Scale periods and Atomic Scale transition periods, when the latter measurements were scaled up in accordance with the discrete self-similar Scale transformation equations of the Self-Similar Cosmological Paradigm (SSCP).

The *second part* of the series, entitled "Discrete Cosmological Self-Similarity and Delta Scuti Variable Stars," demonstrated that the δ Scuti class is analogous to more massive atoms such as carbon, oxygen and nitrogen in excited states characterized by $3 \leq n \leq 6$ and $0 \leq l \leq n-1$. The high degree of heterogeneity in this class of variable stars prevented a straightforward period comparison test for large samples of δ Scuti stars and C/N/O atoms, as was possible in the RR Lyrae case. However, a specific δ Scuti star undergoing pure radial double-mode oscillations was shown to be quantitatively self-similar to its Atomic Scale counterpart: a $^{12}$C atom undergoing a radial mode transition between the $n = 5$ and $n = 4$ energy levels.

In this *final part* of the 3-part study devoted to the discrete cosmological self-similarity of variable stars, we turn our attention to ZZ Ceti variable stars. The complex multi-mode oscillation behavior of ZZ Cetis, their substantial mass range and their broad range of oscillation periods pose a considerable challenge to analysis in terms of the SSCP.

The SSCP emphasizes nature's intrinsic and well-stratified hierarchical organization, proposing that the hierarchy is divided into discrete cosmological Scales, of which we can currently observe the Atomic, Stellar and Galactic Scales. The SSCP also proposes that the discrete Scales are rigorously self-similar to one another, such that for each class of fundamental particle, composite system or physical phenomenon on any given Scale there are self-similar analogues on all other Scales. At present the number of Scales cannot be known, but for reasons of natural philosophy it is tentatively proposed that there are a denumerably infinite number of Scales, ordered in terms of their intrinsic ranges of space, time and mass scales. The spatial (R), temporal (T) and mass (M) parameters of discrete self-similar analogues on neighboring Scales $\Psi$ and $\Psi$-1 are related by the following set of discrete self-similar Scale transformation equations.

$$R_\Psi = \Lambda R_{\Psi-1} \quad , \quad (1)$$

$$T_\Psi = \Lambda T_{\Psi-1} \quad \text{and} \quad (2)$$

$$M_\Psi = \Lambda^D M_{\Psi-1} \quad , \quad (3)$$

where $\Lambda$ and D are empirically determined dimensionless scale factors equal to $5.2 \times 10^{17}$ and 3.174, respectively. The value of $\Lambda^D$ is $1.70 \times 10^{56}$. The symbol $\Psi$ is used as an index to distinguish different Scales, such that

$$\Psi = \{\ldots, -2, -1, 0, +1, +2, \ldots\} \quad , \quad (4)$$

and the Stellar Scale is usually assigned $\Psi = 0$. Thus, Atomic Scale systems and phenomena are designated by $\Psi = -1$ and Galactic Scale systems and phenomena are assigned $\Psi = +1$. The fundamental self-similarity of the SSCP and the recursive character of the discrete scaling equations suggest that nature is an infinite discrete fractal, in terms of its morphology, kinematics and dynamics. A more detailed review of the SSCP can be found in two published papers[1,2] or at the author's website.[3]

There are two main goals to be achieved in the present study. Firstly, using information about the general properties of ZZ Ceti stars, we hope to identify the Atomic Scale analogues of ZZ Cetis and to understand the physical states and processes associated with ZZ Ceti phenomena. Secondly, we attempt to define and pursue empirical tests that will help in convincing us that we have correctly analyzed a unique manifestation of discrete cosmological self-similarity between Atomic and Stellar Scale systems.

2. PROPERTIES OF ZZ CETI VARIABLE STARS

ZZ Ceti stars are variable white dwarf stars, also known as DAV stars.[4] Their all-important mass range[5] is roughly from 0.53 $M_\odot$ to at least 1.10 $M_\odot$. Typical radii for white dwarf stars are on the order of $10^9$ cm, or $10^4$ km. Most of their mass is thought to be in a core object composed of He, C, O, and heavier atoms.[6] Surrounding the core is a thin envelope of pure H whose mass has a canonical value[7] of $\approx 10^{-4}$ $M_{star}$, or roughly 6 x $10^{-5}$ $M_\odot$ to 8.7 x $10^{-5}$ $M_\odot$ depending on modeling assumptions.[4,5,8] It is thought that the ZZ Ceti pulsations "are driven in the partial hydrogen ionization zone."[4] ZZ Ceti variables are multi-periodic, non-radial pulsators with an *average* of 3 independent low-amplitude oscillation modes per star, and ≥ 20 modes in one star is not uncommon. ZZ Ceti periods can be amazingly stable[9] or they can vary significantly within a fraction of a day. The period range for ZZ Cetis is about 100 – 1200 sec and a histogram of observed oscillation periods for a sample[6] of 82 ZZ Cetis is shown in Figure 1.

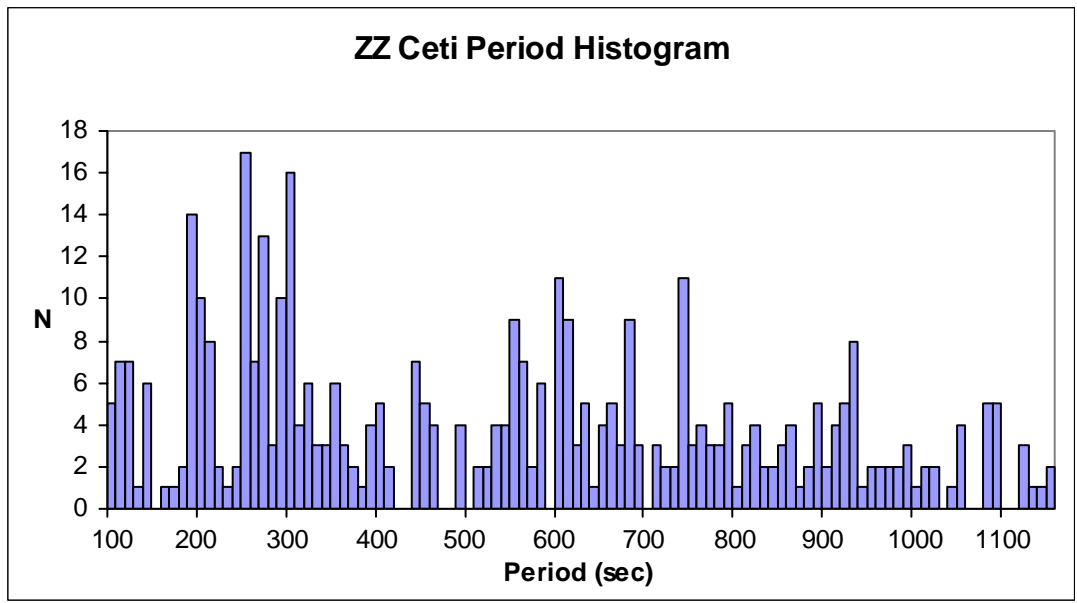

**Figure 1.** Distribution of periods for a combined sample of 82 ZZ Ceti variables from a paper by Mukadam et al.[6] Data include all independent periods measured for these stars; some stars are observed multiple times by one or more research groups.

Because ZZ Ceti pulsation behavior is typically multi-periodic, non-radial, non-sinusoidal and low-amplitude, their unprocessed frequency spectra can include independent pulsations, various harmonics, combination frequencies, difference frequencies and aliases. In spite of many factors that might serve to mask discretization in ZZ Ceti frequency spectra, Figure 1 is certainly suggestive of a heterogeneous collection of relatively discrete periods. Bergeron et al[10] comment that "these pulsators show typically only a few excited modes out of the very rich g-mode spectra available to them." Table 1 summarizes the basic properties of ZZ Ceti variables.

**Table 1.** ZZ Ceti Properties

| | |
|---|---|
| **Mass Range** | 0.53 $M_\odot$ to 1.10 $M_\odot$ |
| **Radii** | ~ $10^9$ cm |
| **Pulsation Properties** | Multi-mode, non-radial, low amplitude, non-sinusoidal |
| **Number of Periods** | Average = 3, Maximum ≥ 20 |
| **Period Range** | 100 to 1200 sec |

Several hundred ZZ Ceti stars have been identified to date and since they are relatively faint stars, they are all within 400 parsecs of the Sun[8]. Therefore observed ZZ Cetis are Population I members of the Galaxy's <u>disk</u> population, in contrast to most classical variable stars, which tend to be Population II objects and are more typically found throughout the Galaxy, including its halo and globular clusters.

3. SSCP INTERPRETATION OF ZZ CETI PHENOMENA

Of the three classes of variable stars that have been studied in this series (RR Lyraes, δ Scutis and ZZ Cetis), the ZZ Ceti class poses the biggest challenges to analysis within the context of the SSCP. Even within the context of conventional astrophysics the ZZ Cetis remain relatively enigmatic. Here we must be satisfied with achieving a less comprehensive and quantitative understanding than was possible in the case of RR Lyrae stars. However, in spite of the difficult nature of the subject, we will succeed in constructing and testing a first approximation SSCP assessment of ZZ Ceti phenomena that will serve as the foundation for future refinements in our understanding of the discrete cosmological self-similarity of ZZ Ceti stars.

The key parameter in SSCP investigations of variable stars is the mass range of the relevant systems. The approximate ZZ Ceti mass range is $0.53\ M_\odot \leq M \leq 1.10\ M_\odot$, and Eq. (3) can be used to calculate that the corresponding masses of Atomic Scale analogues range from 4 to 8 atomic mass units (amu). This mass range includes $^4$He, $^6$He, $^6$Li, $^7$Li, $^8$Li, $^7$Be, $^8$Be and $^8$B. Given reasonable assumptions based on observed abundances for various atoms and their isotopes, we can expect $^4$He to be the most common Atomic Scale analogue of ZZ Ceti stars, with a smaller contribution from $^7$Li analogues, and probably only minor or trace contributions from analogues of $^6$He, $^{6,7}$Li, $^{7,8}$Be, $^8$B and heavier atoms.

Unlike RR Lyraes and δ Scutis, which are analogues of atoms in *neutral* states, ZZ Cetis appear to be analogues of *positive ions*. Two physical properties lead to this conclusion. Firstly, the extremely small size of white dwarf stars and the scaling of Eq. (1) require an analogy to small tightly bound wavefunctions that are only found in certain ionic states, such as $He^+$ and more massive ions with few remaining electrons.[1]

Secondly, the unique structure of ZZ Ceti stars, especially the thin pure H outer envelope with a mass on the order of $10^{-4}$ $M_\odot$, also argues for one or two tightly bound electrons, since the mass of the Stellar Scale electron has been determined[11] previously to be ≈ 7 x $10^{-5}$ $M_\odot$. Therefore, the SSCP proposes that the majority of ZZ Ceti stars are Stellar Scale analogues of $^4He^+$, $^7Li^+$ and $^7Li^{++}$.

Having defined the primary Atomic Scale analogue systems and their ionic states, we would now like to explore the relevant range of n and Δn values that correspond to the oscillations observed for ZZ Ceti stars. In order to make progress in this direction we need to adopt a *crucial hypothesis* about the nature of ZZ Ceti oscillations. We will hypothesize that the dominant independent periods of ZZ Cetis (not including harmonics, aliases, or combination/difference periods) correspond to specific allowed transition periods of their Atomic Scale analogues. It is nearly inconceivable that multi-mode low-amplitude ZZ Cetis are actually undergoing discrete energy level transitions, since that would appear to require from 3 to 20 (or more) simultaneous transitions in systems most typically possessing *one* electron! It seems more reasonable and natural to interpret ZZ Ceti stars in terms of Atomic Scale analogues that are highly perturbed systems in a high-energy Galactic disk environment characterized by relatively strong electric, magnetic, pressure, temperature and radiation fields. Within the context of the SSCP, ZZ Cetis in the Galactic disk are analogous to He and Li ions in the equatorial disk of a Stellar Scale ultracompact object such as a neutron star or black hole.[1,2] Although the multi-mode oscillations may not represent full-fledged energy level transitions, our working hypothesis here is that the perturbed ZZ Cetis are "ringing" at discrete transition frequencies that are rigorously self-similar to the corresponding discrete transition periods of their Atomic Scale analogues, but with amplitudes that are below the threshold for actual energy level transitions. This critical hypothesis will be empirically tested below.

The effort to determine relevant n and $\Delta$n values for the Atomic Scale analogues of RR Lyrae and $\delta$ Scuti stars was greatly facilitated by approximate relationships among the radii, periods and principal quantum numbers of Rydberg atoms. However, these helpful relationships are not applicable in the case of ZZ Ceti stars because they do not appear to be analogues of Rydberg atoms, i.e., neutral atoms with the single outermost electron excited to relatively high values of n. To make progress toward n and $\Delta$n estimates it is necessary to take an entirely empirical approach. We proceed by calculating the oscillation periods for all energy level transitions for $^4$He$^+$ ions between $1 \leq n \leq 8$, including $\Delta$n values of $1 \leq \Delta n \leq 6$. Our effort is greatly simplified by the fact that the $^4$He$^+$ ion has a remarkably degenerate energy level structure. This means that to a very good approximation ($\leq$ 1% errors) we can treat the energy levels as discrete single-valued functions of n, i.e., with virtually no level splitting due to other quantum numbers. Table 2 presents the calculations based on data found in a standard source[12] for atomic data, and lists predicted Stellar Scale oscillation periods derived from the $^4$He$^+$ transition periods and Eq. (2).

**Table 2.** He$^+$ Transition Periods and Stellar Scale Analogue Periods

| Transition | Δn | ΔE (cm$^{-1}$) | Observed Atomic Scale Period (sec) | Predicted Stellar Scale Period (sec) |
|---|---|---|---|---|
| 6 ↔ 1 | 5 | 426717.129 | 7.812 x 10$^{-17}$ | 40.6 |
| 4 ↔ 1 | 3 | 411477.158 | 8.101 x 10$^{-17}$ | 42.1 |
| 2 ↔ 1 | 1 | 329179.744 | 1.013 x 10$^{-16}$ | 52.7 |
| | | | | |
| 8 ↔ 2 | 6 | 102871.310 | 3.240 x 10$^{-16}$ | **168.5** |
| 7 ↔ 2 | 5 | 100771.958 | 3.308 x 10$^{-16}$ | **172.0** |
| 6 ↔ 2 | 4 | 97537.385 | 3.417 x 10$^{-16}$ | **177.7** |
| 5 ↔ 2 | 3 | 92172.941 | 3.616 x 10$^{-16}$ | **188.1** |
| 4 ↔ 2 | 2 | 82297.414 | 4.050 x 10$^{-16}$ | **210.6** |
| 3 ↔ 2 | 1 | 60961.198 | 5.468 x 10$^{-16}$ | **284.3** |
| | | | | |
| 8 ↔ 3 | 5 | 41910.112 | 7.954 x 10$^{-16}$ | **413.6** |
| 7 ↔ 3 | 4 | 39810.760 | 8.373 x 10$^{-16}$ | **435.4** |
| 6 ↔ 3 | 3 | 36576.187 | 9.113 x 10$^{-16}$ | **473.9** |
| 5 ↔ 3 | 2 | 31211.743 | 1.068 x 10$^{-15}$ | **555.3** |
| 4 ↔ 3 | 1 | 21336.216 | 1.562 x 10$^{-15}$ | **812.4** |
| | | | | |
| 8 ↔ 4 | 4 | 20573.896 | 1.620 x 10$^{-15}$ | **842.5** |
| 7 ↔ 4 | 3 | 18474.544 | 1.804 x 10$^{-15}$ | **938.2** |
| 6 ↔ 4 | 2 | 15239.971 | 2.187 x 10$^{-15}$ | **1137.4** |
| 5 ↔ 4 | 1 | 9875.529 | 3.375 x 10$^{-15}$ | 1755.2 |
| | | | | |
| 8 ↔ 5 | 3 | 10698.369 | 3.116 x 10$^{-15}$ | 1620.2 |
| 7 ↔ 5 | 2 | 8599.017 | 3.876 x 10$^{-15}$ | 2015.7 |
| 6 ↔ 5 | 1 | 5364.444 | 6.214 x 10$^{-15}$ | 3231.2 |
| | | | | |
| 8 ↔ 6 | 2 | 5333.925 | 6.249 x 10$^{-15}$ | 3249.6 |
| 7 ↔ 6 | 1 | 3234.573 | 1.031 x 10$^{-14}$ | 5358.8 |

We see immediately that three transition period *series* associated n = 2, n = 3 and n = 4 correspond well with the range of periods for ZZ Ceti stars. This is made clear in Figure 2, which superimposes relevant data from Table 1 onto our histogram of ZZ Ceti periods (Fig. 1).

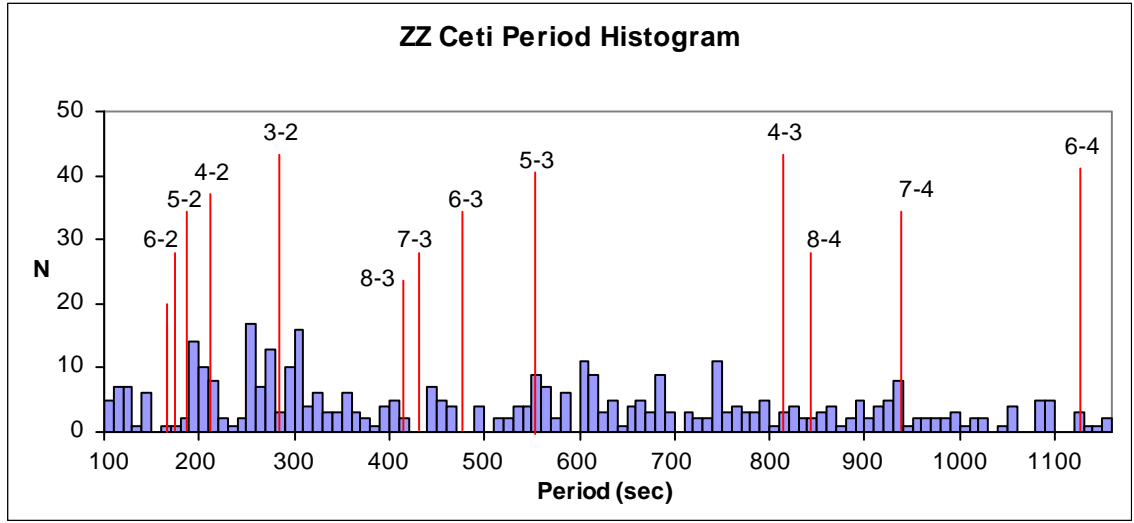

**Figure 2**. The histogram of ZZ Ceti periods from Fig. 1 is overlain with red lines designating the positions of scaled transition periods of $^4$He$^+$ in the 100 to 1200 sec range (see Table 1).

Our largely heuristic SSCP analysis of ZZ Ceti phenomena is summarized in Table 3.

**Table 3.** SSCP Analysis of the ZZ Ceti Class

| Atomic Scale analogues | He$^+$, Li$^+$, Li$^{++}$ (Be$^{+, ++, +++}$) |
|---|---|
| Most likely n values | 2,3,4 |
| n range | $2 \leq n \leq 8$ |
| Δn of oscillations | $1 \leq \Delta n \leq 6$ |
| l range | $1 \leq l \leq 7$ |
| Oscillating at discrete transition periods? | Probably |
| Undergoing energy level transitions? | Probably not |

4. A TEST OF THE SSCP INTERPRETATION

Because the SSCP analysis of ZZ Ceti variable stars is quite speculative and requires the crucial hypothesis that the low-amplitude multi-mode oscillation periods of ZZ Cetis are directly related to the discrete transition periods of their Atomic Scale counterparts, it is highly desirable to test the SSCP analysis in a quantitative manner. A relevant test has been identified and conducted, but before presenting the results it is prudent to briefly review the physical factors that can interfere with a straightforward determination of discrete self-similarity between Atomic Scale and Stellar Scale analogues. These primary physical considerations are listed in Table 4 below.

**Table 4.** Factors Influencing Self-Similarity Tests for Atomic/Stellar Systems

| Ambient Galactic Scale parameters that can cause shifts in stellar oscillation periods | Factors that generate additional or extraneous oscillation periods |
|---|---|
| Electric fields | Isotope heterogeneity |
| Magnetic fields | Analogue heterogeneity |
| Temperature | Singlet, doublet, … multiplicity |
| Pressure | Ionic state heterogeneity |
| Radiation fields | Harmonics, aliases, combin./diff. periods |

The ZZ Ceti sample of Bergeron *et al*[5] has a considerable range of masses, as shown in Table 5. If our preliminary analysis of ZZ Ceti phenomena is on the right track, then the lowest mass stars are $^4He^+$ analogues and their dominant oscillation periods will match up with scaled transition periods for $^4He^+$. On the other hand, the high mass stars from the sample should *not* be $^4He^+$ analogues and their dominant periods should *not* match up well with the scaled $^4He^+$ transition periods. For our $^4He^+$ analogue subset we choose the four lowest mass systems (0.53 $M_\odot \leq M \leq$ 0.55 $M_\odot$) and concern ourselves only with the dominant oscillation period of each star. As shown in Figure 3, the results of this test agree with our expectations: each of the four dominant periods lines up close to one of the scaled transition periods of $^4He^+$.

**Table 5.** Mass[5] and Dominant Periods[6] for a sample of ZZ Ceti Variables

| Star ID | Mass ($M_\odot$) | Dominant Period (sec) |
|---|---|---|
| LP 133 -144 | 0.53 | 209.18 |
| HL Tau 76 | 0.55 | 541 |
| G38 – 29 | 0.55 | 938.0, 910.3 = <924.2> |
| G238 – 53 | 0.55 | 206.2 |
| G30 – 20 | 0.58 | 1068 |
| R548 | 0.59 | 213.1 |
| G117 – B15A | 0.59 | 215.2 |
| BPM 30551 | 0.75 | 741, 744 = <742.5> |
| HE 0532 – 5605 | 0.92 | 688.8 |
| LTT 4816 | 1.10 | 192 |
| G226 – 29 | 0.79 | 109.5 |
| G207 - 9 | 0.83 | 318.0 |

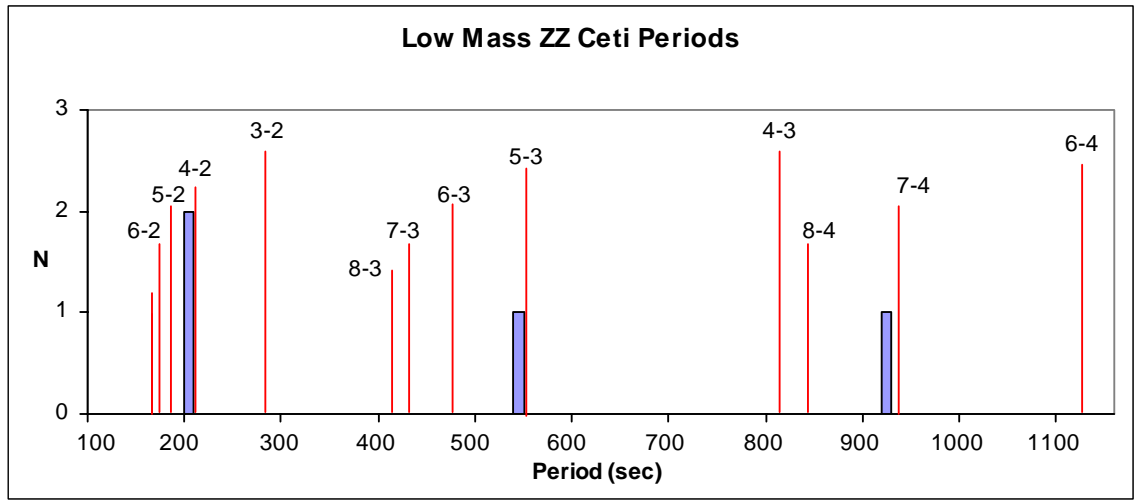

**Figure 3.** Superposition of the dominant periods of the low mass ZZ Ceti subset and the scaled $^4He^+$ transition periods.

It is interesting that in each case the observed period is slightly shorter than the matching predicted period, suggesting the possibility of a small systematic shift such as might be caused by an ambient magnetic field in the Galactic disk. The quantitative data for this test are given in Table 6.

**Table 6.** Quantitative Data Shown in Figure 3

| Star ID | Mass ($M_\odot$) | Dominant Period (sec) | Closest Scaled $^4He^+$ Period (sec) | Error (%) |
|---|---|---|---|---|
| LP 133 – 144 | 0.53 | 209.2 | 210.6 | 0.7 |
| HL Tau 76 | 0.55 | 541 | 555.3 | 2.6 |
| G38 – 29 | 0.55 | <924.2> | 938.2 | 1.5 |
| G238 - 53 | 0.55 | 206.2 | 210.6 | 2.1 |

As a check on the uniqueness of the agreement between the observed periods of the low mass subset and the predicted set of Stellar Scale periods, we can repeat the test for the high mass subset of ZZ Cetis listed in Table 5. Our second subset consists of five stars with masses ranging from 0.75 $M_\odot$ to 1.10 $M_\odot$, and the dominant modes for these stars are compared with the scaled $^4He^+$ transition periods in Figure 4. For this subset of high mass ZZ Cetis we see very little agreement between the observed and predicted periods, as expected.

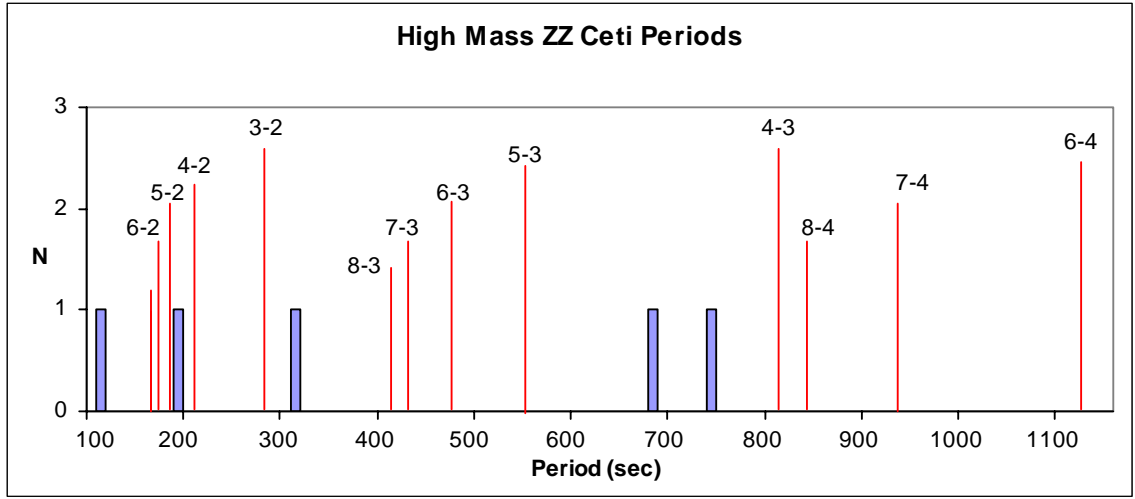

**Figure 4.** Superposition of the dominant periods of the high mass ZZ Ceti subset and the scaled $^4He^+$ transition periods.

This result supports the contention that the agreement found in the low mass ZZ Ceti case was unique rather than coincidental. Quantitative data for this second part of our uniqueness test of the SSCP analysis are presented in Table 7.

**Table 7.** Quantitative Data Shown in Figure 4

| Star ID | Mass ($M_\odot$) | Dominant Period (sec) | Closest Scaled $^4He^+$ Period (sec) | Error (%) |
|---|---|---|---|---|
| BPM 30551 | 0.75 | <742.5> | 812.4 | 8.6 |
| HE 0532-5605 | 0.92 | 688.8 | 812.4 | 15.2 |
| LTT 4816 | 1.10 | 192 | 188.1 | 2.7 |
| G226-29 | 0.79 | 109.5 | 168.5 | 35.0 |
| G207-9 | 0.83 | 318.0 | 284.3 | 11.9 |

This test of the SSCP analysis of ZZ Ceti phenomena constitutes a preliminary falsification test of the discrete cosmological self-similarity between low mass ZZ Ceti stars and $^4He^+$ ions. If the dominant modes of the low mass subset did not match up convincingly with the scaled $^4He^+$ transition periods, then the whole SSCP analysis would have been in serious doubt. The successful matches between observed and predicted periods offer encouragement for further exploration of the discrete cosmological self-similarity of ZZ Ceti phenomena.

5. CONCLUSIONS

In conclusion, the SSCP interprets the overwhelming majority of ZZ Ceti stars as discrete self-similar analogues of $He^+$, $Li^+$ and $Li^{++}$ ions. These systems are highly perturbed and are hypothesized to be oscillating at discrete transition periods, but their low-amplitude and multi-mode characteristics seem to argue that the oscillations are below the threshold for actual energy level transitions.

The helium ion analogues, which should outnumber the lithium ion analogues, have relevant n values of $2 \leq n \leq 8$ and relevant $\Delta n$ values of $1 \leq \Delta n \leq 6$. Preliminary results indicate that the ranges of n and $\Delta n$ are very similar for the lithium ion analogues, but an analysis of this smaller, more complicated, subset of analogues must remain a project for the future. The helium analogues appear to be oscillating at specific sets of transition periods associated with transitions to or from $n = 2$, $n = 3$ and $n = 4$.

A more complete understanding of ZZ Ceti stars will require inclusion of scaled frequency spectra for the more complicated Li ions and heavier ions. Accurate mass and radius data for individual stars would greatly reduce the complexity of the research effort by putting limits on relevant Atomic Scale analogues, their ionic states and their energy levels.

REFERENCES

1. Oldershaw, R.L., *Internat. J. Theor. Phys.* **12**(6), 669-694, 1989

2. Oldershaw, R.L., Internat. J. Theor. Phys. 12(12), 1503-1532, 1989.

3. Oldershaw, R.L., http://www.amherst.edu/~rloldershaw .

4. Kim, A., *et al.*, *Mem. Soc. Astron. Ital.*, 2004 (see also preprint: arXiv:astro-ph/0507490 v1, available at http://www.arxiv.org , 2005).

5. Bergeron, P., *et al.*, *Astrophys. J.* **600**, 404-408, 2004.

6. Mukadam, A.S., *et al.*, *Astrophys. J.* **635**(2), 1239-1262, 2005.

7. Althaus, L.G., *et al.*, *Astron. & Astrophys.* **440**, L1-L4, 2005.

8. Kepler, S.O., *et al.*, *Astrophys. J.* **534**, L185-L188, 2000.

9. Kepler, S.O., *et al.*, *Astrophys. J.* **634**, 1311-1318, 2005.

10. Bergeron, P., *et al.*, *Astrophys. J.* **449**, 258-279, 1995.

11. Oldershaw, R.L., *Astrophys. J.* **322**, 34-36, 1987.

12. Bashkin, S. and Stoner, Jr., J.O., *Atomic Energy Levels and Grotrian Diagrams (Vol. I. Hydrogen I – Phosphorus XV)*, North-Holland Pub. Co., Amsterdam; American Elsevier Pub. Co., Inc., New York, 1975.